\documentclass[aip,apl,numerical,reprint]{revtex4-1}
\usepackage{graphicx}%
\usepackage{bm}%
\usepackage{amssymb}
\usepackage[amssymb]{SIunits}
\usepackage[version=3]{mhchem}
\usepackage{hyperref}
\usepackage{textcomp}
\usepackage{color}

\newcommand{\UnivNEELAddress}{Univ. Grenoble Alpes, Inst. NEEL, 
F-38042 Grenoble, France}
\newcommand{\CNRSNEELAddress}{CNRS, Inst. NEEL, F-38042 Grenoble, France}
\newcommand{\CEAAddress}{CEA, INAC, F-38054 Grenoble, France}

\begin{document}

\title{Cathodoluminescence of stacking fault bound excitons for local
   probing of the exciton diffusion length in single GaN nanowires }

\author{Gilles Nogues}
\email{gilles.nogues@neel.cnrs.fr}
\affiliation{\UnivNEELAddress}
\affiliation{\CNRSNEELAddress}

\author{Thomas Auzelle}
\affiliation{\CEAAddress}

\author{Martien Den Hertog}
\affiliation{\UnivNEELAddress}
\affiliation{\CNRSNEELAddress}

\author{Bruno Gayral}
\affiliation{\CEAAddress}

\author{Bruno Daudin}
\affiliation{\CEAAddress}

\date{\today{}}


\begin{abstract}  We perform correlated studies of individual \ce{GaN} nanowires 
in scanning electron microscopy combined to low temperature cathodoluminescence, 
microphotoluminescence and scanning transmission electron microscopy. We show 
that some nanowires exhibit well localized regions emitting light at the energy 
of a stacking fault bound exciton (\unit{3.42}{\electronvolt}) and are able to 
observe the presence of a single stacking fault in these regions. Precise 
measurements of the cathodoluminescence signal in the vicinity of the stacking 
fault gives access to the exciton diffusion length near this location. 
\end{abstract}

\maketitle

Carrier diffusion length is a key quantity in optoelectronics, as it notably 
plays a significant role in the competition between radiative and non-radiative 
processes, especially in materials with large densities of defects. Concerning 
nitride semiconductors, various studies have established that carrier diffusion 
length in \ce{InGaN} and \ce{GaN} bidimensional (2D) layers are rather small, in 
the range of 
50-\unit{250}{\nano\meter}\cite{SpeckRosner_rolethreadingdislocations_99, 
RosnerErikson_Correlationcathodoluminescenceinhomogeneity_97, 
RosnerSpeck_Cathodoluminescencemappingepitaxial_99, 
SugaharaSakai_DirectEvidencethat_98, 
SondereggerGaniere_Highspatialresolution_06, 
LiuoliaNakamura_Dynamicspolarizedphotoluminescence_10, 
ChichibuNakamura_Spatiallyresolvedcathodoluminescence_97}. Such lengths are 
smaller than the typical distance between 
dislocations\cite{Godlewskiet_Diffusionlengthcarriers_04}, one among other 
possible reasons for the surprising efficiency of radiative recombination in 
\ce{InGaN} quantum wells in spite of the high density of defects in current 
heterostructures. With the aim of further efficiency improvement, a current 
trend in nitride optoelectronics research is to explore the potential of 
nanowires (NWs) as building-blocks for light emission or absorption devices. 
This approach has already led to the realization of NW heterostructure-based 
light-emitting diodes (LEDs)\cite{KikuchiKishino_InGaN/GaNMultipleQuantum_04, 
TourbotDaudin_Structuralandoptical_11, 
BavencoveDang_Submicrometreresolvedoptical_11, GuoOoi_InGaN/GaNdiskin_11}. In 
the case of these pioneering works, the small diameter of NWs raises the issue 
of the influence of size effects on carrier diffusion.  More generally, the 
issue of carrier diffusion length in GaN NWs is poorly documented to date. 
Recent works\cite{SchlagerSanford_Steady-stateandtime-resolved_08, 
ParkTaylor_GaNnanorodsgrown_11, ChenYang_Lightemittingdevice_13} suggest that 
recombinations are faster in \ce{InGaN} nanorods than in 2D layers, which is 
correlated to a smaller carrier diffusion length and has been tentatively 
assigned to possible surface damage. In this letter we address the issue of 
exciton diffusion length in single \ce{GaN} NWs with a diameter in the range of 
\unit{100}{\nano\meter}. For this purpose, low-temperature cathodoluminescence 
(CL) experiments have been performed on individual NWs in correlation with 
microphotoluminescence (\textmu PL) and scanning transmission electron 
microscopy (STEM) studies that reveal the presence of I1 basal stacking faults 
(SFs). It is well known from 2D-layer studies that such SFs are radiative 
recombination centers\cite{SalviatiStrunk_CathodoluminescenceandTransmission_99, 
LiuKhan_Luminescencefromstacking_05, 
CorfdirDeveaud-Pledran_Excitonlocalizationbasal_09}. Previous TEM/CL studies on 
\ce{GaN} epilayers have already evidenced the correlation between the presence 
of different types of SFs with well identified emission 
peaks\cite{SalviatiStrunk_CathodoluminescenceandTransmission_99, 
LiuKhan_Luminescencefromstacking_05}. The density of structural defects was 
relatively large in those samples. In the case of our NWs, we show that emission 
peaks are linked to the presence of a single defect, acting as a quasi-punctual 
recombination center. By varying the distance between the SF and the the CL 
excitation spot we investigate the diffusion length along the NW axis.


The nanowires are grown by Plasma-assisted Molecular Beam Epitaxy (MBE) on a 
2-inch \ce{Si}(111) substrate. Desoxydation of the silicon is done in situ by 
annealing at \unit{950}{\celsius} until the clear appearance of the 7$\times$7 
surface reconstruction at \unit{820}{\celsius}. The growth temperature, set at 
\unit{820}{\celsius}, is determined by the measurement of the corresponding 
\ce{Ga} desorption time \cite{LandreDaudin_Plasmaassistedmolecular_08, 
MataDaudin_NucleationGaNnanowires_11}. A thin \ce{AlN} buffer layer 
(2-\unit{3}{\nano\meter} thick) is grown directly onto the silicon. Used as a 
seed layer for the \ce{GaN} nanowires, it decreases the tilt of the NWs relative 
to the normal to the surface\cite{SongmuangDaudin_Fromnucleationto_07}. This 
helps to obtain well separated nanowires, even for extensive growth time. NWs 
were grown for \unit{18}{\hour} in N-rich conditions  (Ga/N ratio of 
0.3)\cite{Sanchez-GarciaBeresford_effectIII/Vratio_98}. They are 
$L=$\unit{3}{\micro\meter} long and their diameters range from 50 to 
\unit{100}{\nano\meter}. 

\begin{figure}[hb]
 \centering
\includegraphics[width=.90\columnwidth]{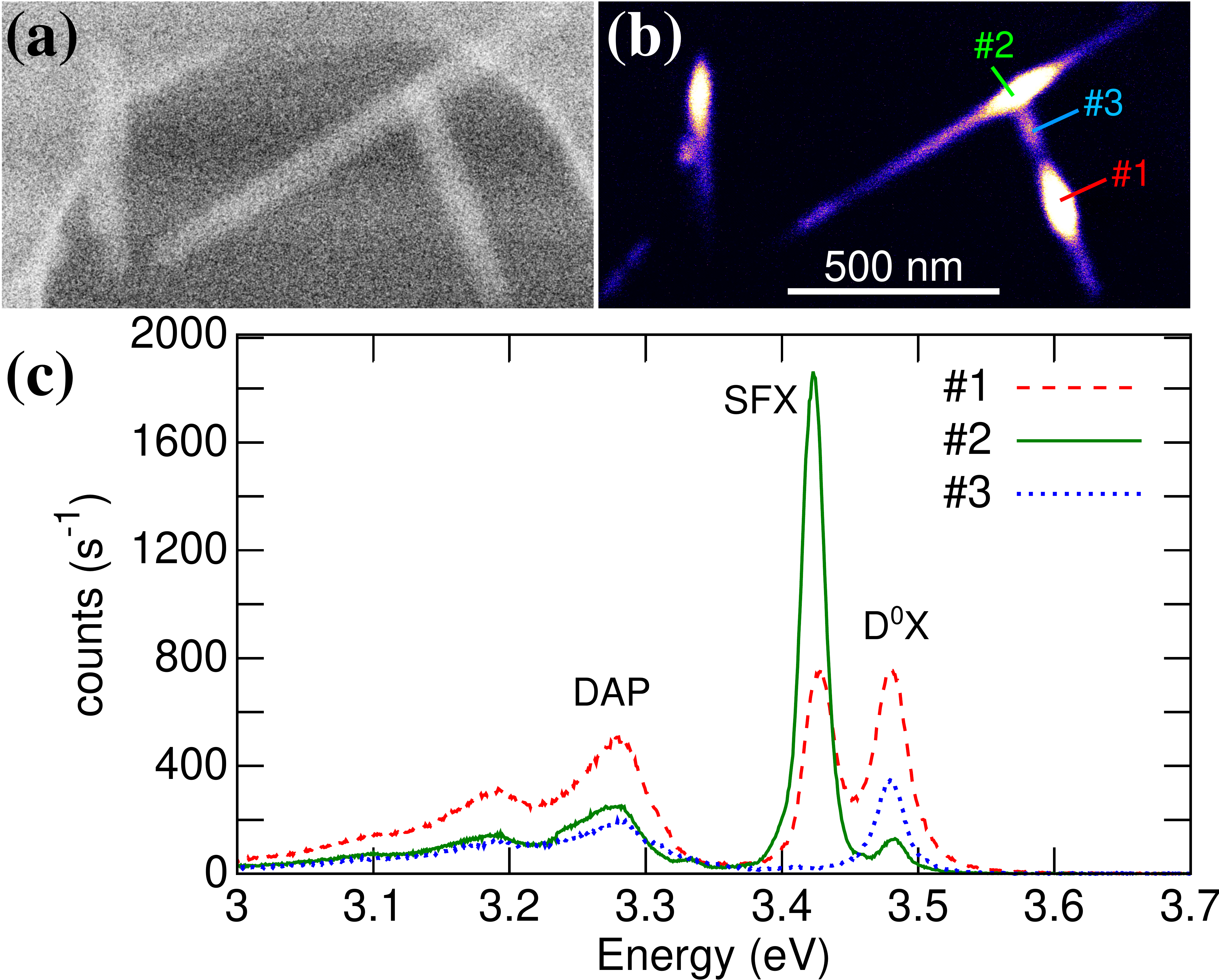}
  \caption{(a) SEM image of 3 dispersed NWs. (b) Corresponding image of the CL 
signal between \unit{3.40}{\electronvolt} and \unit{3.50}{\electronvolt}. (c) CL 
spectra at \unit{5}{\kelvin} with a spot excitation at points \#1 to \#3 shown 
on (b).
  }\label{fig:spectra}
\end{figure}

As grown NWs are mechanically dispersed onto 2 kinds of substrates. Substrate S1 
is used for CL and \textmu PL studies. It is made of a \ce{Si} wafer. Localization 
marks are patterned by standard deep UV (DUV) optical lithography followed 
by reactive ion etching (RIE), using the resist as a mask. It is then sputtered 
with a \unit{100}{\nano\meter}-thick \ce{Al} layer and a 
\unit{20}{\nano\meter}-thick \ce{SiO2} layer. For combined TEM and CL studies, 
substrate S2 consists of a home-made \unit{35}{\nano\meter} thick \ce{Si3N4} 
membrane with a window size of \unit{90}{\micro\meter}. Arrays of nitride 
membranes are fabricated starting from a \unit{200}{\micro\meter} thick \ce{Si} 
(100) wafer with a layer of low stress \ce{Si3N4} of \unit{35}{\nano\meter} on 
top of a \unit{240}{\nano\meter} thick \ce{SiO2} layer on each side. The 
fabrication procedure is described in 
Ref.~\cite{HertogMonroy_CorrelationPolarityand_12}, with the difference that for 
these membranes the \ce{SiO2} layer is removed below the \ce{Si3N4} layer by RIE 
and a sequential \ce{KOH} etch. Optical DUV lithography and electron beam 
metallization are used to pattern \ce{Ti}-\ce{Au} markers on the membranes to 
locate the same NW in different experiments.

CL images are taken at \unit{5}{\kelvin} in a FEI scanning electron microscope 
(SEM) and analyzed through a \unit{45}{\centi\meter} spectrometer and 
UV-optimized grating with 600 grooves/\milli\meter. Figures 
\ref{fig:spectra}(a-b) present simultaneously recorded SEM and CL images of 3 
NWs on substrate S1. The color scale of the CL image is set in order to show 
that light comes from the whole length of the NWs. Nevertheless, a strong 
emission arises from 3 bright spots which saturate the image. Figure 
\ref{fig:spectra}(c) presents three spectra with a fixed electron beam 
excitation at three locations (\#1 to \#3) shown on image \ref{fig:spectra}(b). 
The spectrum \#3 is representative of the emission from the whole length and is 
similar to the ensemble PL measurements on the as-grown sample. One observes a 
peak at \unit{3.48}{\electronvolt} that corresponds to the emission from neutral 
donor bound exciton (D$^0$X) and a peak at \unit{3.28}{\electronvolt} with its 
phonon replica at lower energy that are attibuted to donor-acceptor pairs (DAP). 
Spectra \#2 and \#3 show that the bright spots are associated to an extra 
emission peak at \unit{3.42}{\electronvolt} attributed to the emission 
from excitons trapped by stacking 
faults (SFX)\cite{RebaneAlbrecht_StackingFaultsas_97, PhysRevB.57.R15052, 
SalviatiStrunk_CathodoluminescenceandTransmission_99}. CL observations allow us 
to locate well isolated NWs with SFX emission that are further studied by 
microphotoluminescence. \textmu PL spectra are similar to 
Fig.~\ref{fig:spectra}(c) with a magnitude of the \unit{3.42}{\electronvolt} 
peak slightly smaller or comparable to the D$^0$X one. This is well understood 
by considering that the laser excitation spot is larger than the NW and that the 
\textmu PL spectrum integrates light emitted over its whole volume. The 
linewidth of the D$^0$X peak varies significantly from one NW to the other 
between 2 and \unit{20}{\milli\electronvolt}, whereas the 
\unit{3.42}{\electronvolt} peak linewidth is always in the 
10-\unit{20}{\milli\electronvolt} range. The larger linewidth of the 
\unit{3.42}{\electronvolt} peak has been reported 
before\cite{BrandtRiechert_SubmeVlinewidth_10} and is still of unknown origin.

\begin{figure}
 \centering
  \includegraphics[width=\columnwidth]{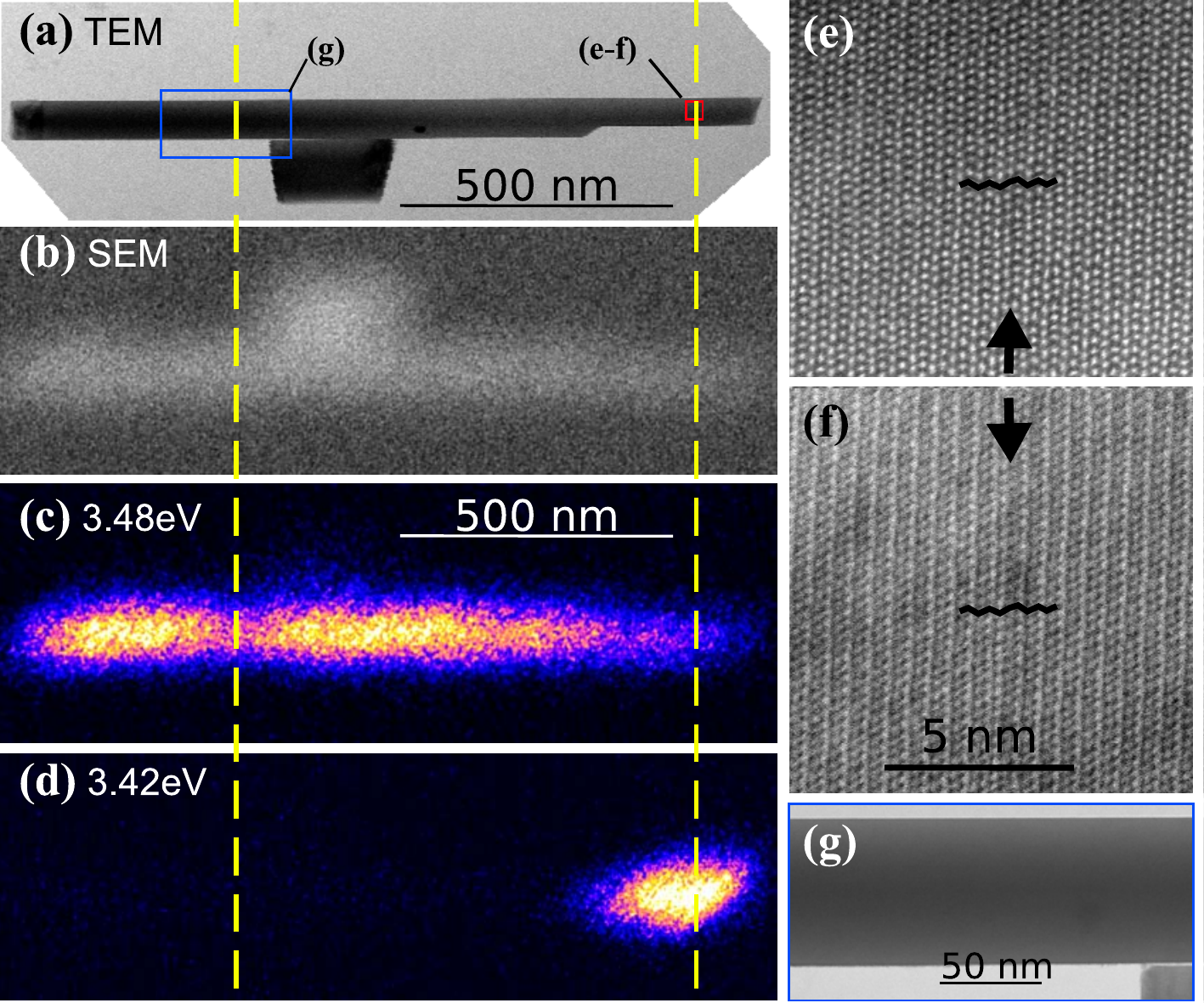}
  \caption{Same NW observed in BF STEM (a), SEM (b), CL at 
\unit{3.48}{\electronvolt} (c) and \unit{3.42}{\electronvolt} (d). (e) HAADF 
STEM  and (f) BF STEM images of the region marked by a red rectangle in (a). (g) 
BF STEM zoom on the blue rectangle in (a).}
  \label{fig:SFplusCL}
\end{figure}

Further evidence of the correlation between the emission at 
\unit{3.42}{\electronvolt} and the presence of a SF is given by joint CL 
and STEM studies of the same NW on a S2 substrate. A well isolated NW is first 
identified in low-temperature CL [Figs.~\ref{fig:SFplusCL}(b-d)]. It displays a 
localized emission at \unit{3.42}{\electronvolt} near one of its tips. The D$^0$X 
signal is more homogeneous, except for two areas where it is 
quenched. One of them corresponds to the place where the 
\unit{3.42}{\electronvolt} emission occurs, the other one is in the middle of 
the NW.  In Figure \ref{fig:SFplusCL}(a) a bright field scanning TEM (BF STEM) 
image of the same NW is shown oriented along the [2-1-10] direction. We note 
that between the two observations, the NW remained at the same location but 
rotated onto itself. It explains why the small piece of \ce{GaN} attached to the 
NW is at different positions between Figs.~\ref{fig:SFplusCL}(a) and (b). The 
region marked by the red rectangle in Fig.~\ref{fig:SFplusCL}(a) is shown at 
higher magnification in BF STEM [Fig.~\ref{fig:SFplusCL}(f)] and high angular 
dark field (HAADF) STEM images [Fig.~\ref{fig:SFplusCL}(e)] which reveal the 
presence of a stacking fault. For sake of clarity it is indicated by an arrow in 
both images, as well as a line following the stacking of the \ce{Ga} columns in 
the 2H hexagonal wurtzite structure (ABABA.. stacking), that is disrupted by the 
insertion of one C plane (..AB\textit{ABC}AC.. stacking) characteristic of a 
cubic zinc blende phase. Comparison of the CL and STEM data clearly shows that 
the 3.42 eV emission exhibits an excellent spatial correlation with the location 
of the SF. On the other hand Fig.~\ref{fig:SFplusCL}(g) shows a higher 
magnification BF STEM image of the region indicated with the blue rectangle in 
Fig~\ref{fig:SFplusCL}(a), that is correlated with a quenching of D$^0$X CL 
emission. In this region we observe no feature indicating the presence of 
crystal defects. A continuous smooth contrast over this region was confirmed by 
high resolution STEM, as for the rest of the NW (not shown here). 
\begin{figure*}
 \centering
 \includegraphics[width=\textwidth]{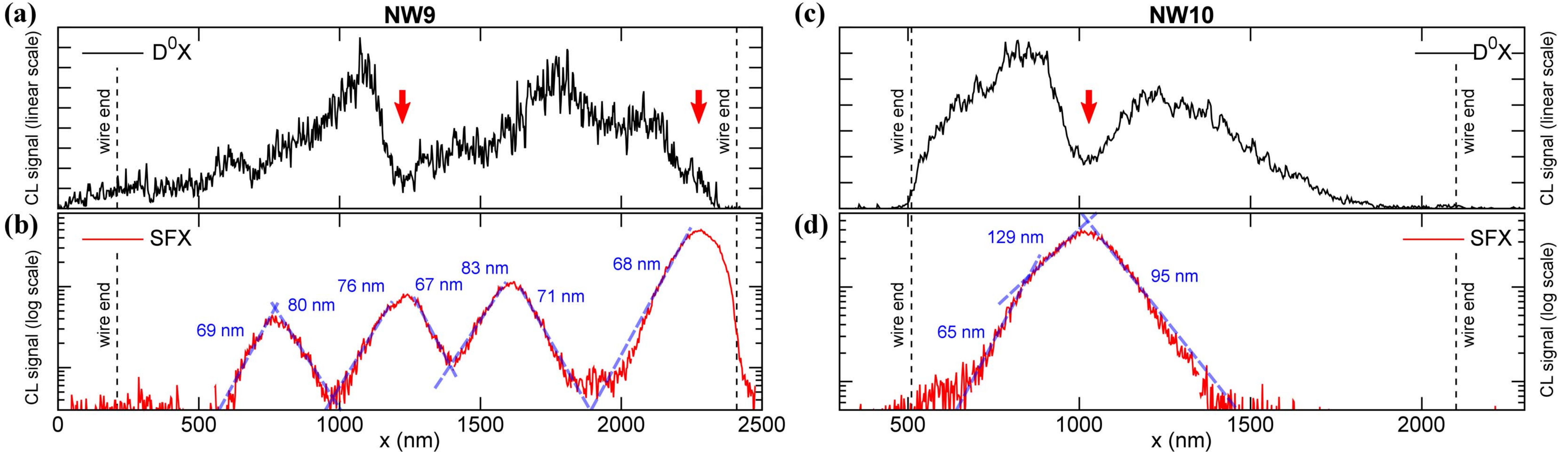}
 \caption{CL intensity profiles along the longitudinal coordinate $x$ for two 
nanowires NW9 (a-b) and NW10 (c-d). (a) and (c): D$^0$X signal in linear scale. 
Arrows mark localized areas where it is unambiguously quenched by the presence 
of a SF. (b) and (d): SFX signal in log scale. Dashed lines are exponential fits 
with characterisctic length $L_{\rm NW}$ indicated next to each line.}
 \label{fig:difflengthfit}
\end{figure*}

The profile of the CL signal along the NW longitudinal axis at 
\unit{3.42}{\electronvolt} is therefore related to the probability for an 
exciton to be trapped by a single SF acting as a radiative recombination center. 
We define $n_{\rm  FX}(x,t)$ (resp. $n_{\rm D^0X}$ and $n_{\rm SFX}$) as the 
linear density of free excitons (FX) (resp. neutral donor and SF bound excitons) 
between $x$ and $x+dx$. Extending the model of Corfdir et 
al.\cite{CorfdirDeveaud-Pledran_Excitonlocalizationbasal_09}, and using the same 
assumptions (absence of non-radiative decay channels and detrapping processes of 
bound excitons), one writes:

\begin{eqnarray}\label{eq:diffusionFX}
 \dfrac{\partial n_{\rm  FX}}{\partial t} & = & D\dfrac{\partial^2 n_{\rm  FX}}{\partial 
x^2} + p(x) - \left( \dfrac{1}{\tau_{\rm r,FX}} + \dfrac{4s}{d} \right) n_{\rm  FX}  \\ 
    & & - \left[ \dfrac{1}{\tau_{\rm FX\rightarrow D^0X}} +\dfrac{1}{\tau_{\rm FX\rightarrow DAP}} + \gamma_{\rm FX\rightarrow \nonumber
SFX}(x)\right] n_{\rm  FX},
\end{eqnarray}

where $D$ is the diffusion constant of the FX in the material. $p(x)$ represents 
the excitation term. It is assumed constant over the wire length in \textmu PL. 
On the contrary it is point like in the case of CL. The FX population can decay 
either radiatively, with time $\tau_{\rm r,FX}$, or through surface effect, $s$ 
being the surface recombination velocity and $d$ the NW 
diameter\cite{Shockley_electronandHoles50}. The last line of 
Eq.~\ref{eq:diffusionFX} corresponds to the trapping of free excitons by neutral 
donors (characteristic time $\tau_{\rm FX\rightarrow D^0X}$), DAPs 
($\tau_{\rm FX\rightarrow DAP}$) or the SF. In the latter case, $\gamma_{\rm 
FX\rightarrow SFX}(x)$ has a non zero value $1/\tau^0_{FX\rightarrow SFX}$ only 
close to the SF, with a characteristic extension $w_{\rm SF}$. In 
the same manner one has:

\begin{eqnarray}
 \dfrac{\partial n_{\rm D^0X}}{\partial t} & = & \dfrac{n_{\rm  FX}}{\tau_{\rm FX\rightarrow 
D^0X}} - \dfrac{n_{\rm D^0X}}{\tau_{\rm r,D^0X}}, \label{eq:diffusionD0X} \\
 \dfrac{\partial n_{\rm SFX}}{\partial t} & = & \gamma_{\rm FX\rightarrow SFX}(x)n_{\rm  FX} 
- \dfrac{n_{\rm SFX}}{\tau_{\rm r,SFX}} \label{eq:diffusionSFX},
\end{eqnarray}

where $\tau_{\rm r,D^0X}$ (resp. $\tau_{\rm r,SFX}$) is the radiative decay time 
of the neutral donor (resp. stacking fault) bound exciton. The total SFX (resp. 
D$^0$X) fluorescence signal is $\int_{-L/2}^{L/2} n_{\rm SFX}/\tau_{\rm r,SFX}dx$ 
(resp. $\int n_{\rm D^0X}/\tau_{\rm r,D^0X}$). \textmu PL observations show that 
their stationnary values have the same amplitude. Hence the SF capture 
rate is dramatically larger than the one of the neutral donors, in agreement 
with the results of 
Ref.~\cite{CorfdirDeveaud-Pledran_Excitonlocalizationbasal_09}. From 
Eqs.~\ref{eq:diffusionD0X} and ~\ref{eq:diffusionSFX}, one infers $\tau^0_{\rm 
FX\rightarrow SFX}\simeq\tau_{\rm FX\rightarrow D^0X}*w_{\rm SF}/L$. Using a SF 
capture range\cite{RebaneAlbrecht_StackingFaultsas_97} $w_{\rm 
SF}=$\unit{3}{\nano\meter}, $1/\tau^0_{FX\rightarrow SFX}$ is 300 times larger 
than the capture rate by neutral donors.

We now assume a single NW extending from $-L/2$ to $L/2$ and containing a single 
SF at $x=0$. It is excited at position $x_p$ by the electron beam, $p(x)=p_0 
\delta(x-x_p)$. Outside the SF capture range, the stationnary solution to 
Eq.~\ref{eq:diffusionFX} is $n_{\rm  FX}(x)=n_0 \exp (-|x-x_p|/L_{\rm NW})$, where 
$L_{\rm NW}=\sqrt{D \tau_{\rm eff}}$ is the diffusion length in the NW, with:
\begin{equation}
 \dfrac{1}{\tau_{\rm eff}} = \dfrac{1}{\tau_{\rm r,FX}} + \dfrac{4s}{d} + 
\dfrac{1}{\tau_{\rm FX\rightarrow D^0X}}+\dfrac{1}{\tau_{\rm FX\rightarrow DAP}}.\label{eq:teffdecay}
\end{equation}

Time-resolved photoluminescence experiments in ensemble of \ce{GaN} nanorods 
reported $\tau_{\rm r,FX}\simeq$\unit{100}{\pico\second}~\cite{ 
HarrisAkasaki_Excitonlifetimesin_95, LahourcadeRuterana_Gakineticsin_08} and 
$s=$\unit{2.7\ \power{10}{4}}{\centi\meter\per\second}~\cite{ 
SchlagerSanford_Steady-stateandtime-resolved_08, 
ParkTaylor_GaNnanorodsgrown_11}. In the set of NWs we have studied 
$d=$110$\pm$\unit{10}{\nano\meter}. Hence the two first terms of 
Eq.~\ref{eq:teffdecay} contribute with the same magnitude to the effective decay 
rate of the FX. The capture rate by neutral donors or DAPs depends on their 
density. For pure undoped thick layers of \ce{GaN},  CL experiments report 
exciton diffusion lengths in the vicinity of threading dislocations ranging from 
$L_{2D}$=81\cite{PaucDrouin_Carrierrecombinationnear_06} to 
\unit{201}{\nano\meter}\cite{InoYamamoto_DIffusionGaNCL_08}. 
In our NW sample, the absence of a peak at the FX energy and the presence of 
D$^0$X and DAP peaks in the spectra let us infer that the capture rate 
$1/\tau_{\rm FX\rightarrow D^0X} + 1/\tau_{\rm FX\rightarrow DAP}$ is larger 
than the radiative or surface decay terms.

Close to a SF, the effective decay rate $1/\tau_{\rm eff}$ is dominated by the 
SF capture rate $1/\tau^0_{\rm FX\rightarrow SFX}$, which is $L/w_{\rm SF}$ 
larger than the D$^0$X capture rate. The corresponding diffusion length is 
therefore divided by a factor $\sqrt{L/w_{\rm SF}}$. It is in the 
\unit{10}{\nano\meter} range, of the same order of magnitude as $w_{\rm SF}$. 
This means that a large fraction of incoming excitons is trapped by the 
SF and later converted into a photon. The SF fluorescence signal is therefore 
proportional to $n_{\rm  FX}(x=0)=n_0 \exp (-|x_p|/L_{\rm NW})$.  


Figures \ref{fig:difflengthfit}(b) and (d) plot the CL signal in logscale as a 
function of the longitudinal wire coordinate $x$ for two different NWs on a S1 
substrate. The peaks are well fitted by convoluting an exponential decay with a 
Gaussian function of width $w_p$=30$\pm$\unit{10}{\nano\meter} for electron 
acceleration voltages $V\ge$\unit{10}{\kilo\volt}. This is larger than $w_{\rm 
SF}$ and rather reflects the size of the excitation region by the electron beam. 
For $V=$\unit{5}{\kilo\volt}, $w_p$ is degraded to \unit{70}{\nano\meter}. We 
attribute this effect to  backscattered electrons in the NW at lower energy. At 
distances larger than $w_p$ from the peak center, the tails on each side are 
well fitted by exponential functions with characteristic length $L_{\rm NW}$ 
(dashed lines). Average of $L_{\rm NW}$ measured for an ensemble of 12 NWs gives 
$L_{\rm NW}$=68$\pm$\unit{16}{\nano\meter}$<L_{2D}$. Based on the previous 
discussion, it confirms that the capture by of neutral donors or DAP dominates 
the FX effective decay rate. We also observe a large dispersion on the values of 
$L_{\rm NW}$, with sometimes abrupt changes like around 
$x=$\unit{800}{\nano\meter} on Fig.~\ref{fig:difflengthfit}(d). This is evidence 
of long range variations of $1/\tau_{\rm eff}$. The latter can arise from 
changes of the density of neutral donors or DAPs or the presence of an extra non 
radiative decay term  $-n_{\rm FX}/\tau_{\rm nr, FX}(x)$ in 
Eq.~\ref{eq:diffusionFX} due to other kind of impurities.




Further evidence of long range variations of the NW parameters is given by the 
CL signal at the D$^0$X emission energy [Figures \ref{fig:difflengthfit}(a) and 
(c)]. Its amplitude variations can only be accounted for if one assumes that 
$\tau_{\rm FX\rightarrow D^0X}$ varies along the wire length or if one adds an 
extra non radiative decay term $-n_{\rm D^0X}/\tau_{\rm nr, D^0X}(x)$ to 
Eq.~\ref{eq:diffusionD0X} due to the trapping of the D$^0$X by other defects. 
For example, a SF can capture excitons from the neighboring neutral 
donors~\cite{CorfdirDeveaud-Pledran_Excitonlocalizationbasal_09} resulting in a 
localized quenching area [arrows on Figs.~\ref{fig:difflengthfit}(a,c)]. 
However, we also  observe extended NW areas where a partial quenching of the 
signal occurs. This can only happen if the previously introduced capture or 
decay rates experience long range variations and are not just point-like as for 
the SF defects. We observe no clear correlation between the 
D$^0$X signal magnitude and $L_{NW}$. This means that $1/\tau_{\rm eff}$ is 
determined by other parameters like DAP capture rate or non radiative decay.


In conclusion, we are able to observe single SFs present in individual NWs with 
different optical and structural techniques. The CL profile in the vicinity of a 
SF is well understood by a simple diffusion model of the exciton.  

\begin{acknowledgments}
This work was performed in the CEA/CNRS joint team ``Nanophysique and 
semiconductors'' of Institut N\'eel and INAC, and in the team ``Structure and 
properties of materials. Extreme conditions'' of Institut N\'eel. We acknowledge 
help from the technical support teams of Institut N\'eel: ``Optics and 
microscopies'' (Fabrice Donatini) and ``Nanofab'' (Bruno Fernandez) and 
benefited from the access to the technological platform NanoCarac of 
CEA-Minatech. We acknowledge financial support from ANR programs JCJC (project 
COSMOS, ANR-12-JS10-0002) and P2N (project FIDEL, ANR-11-NANO-0029).
\end{acknowledgments}

%

\end{document}